\documentclass[10pt,letterpaper]{article}
\usepackage[pdftex]{graphicx}
\usepackage{opex3}
\usepackage{color}
\usepackage{cite}
\usepackage{amsmath,amssymb}
\usepackage{ulem}

\begin{document}

%%%%%%%%%%%%%%%%%% title page information %%%%%%%%%%%%%%%%%%
\title{Coherent control of group index and magneto-optical anisotropy in a multilevel atomic vapor}

\author{Andreas Lampis, Robert Culver, Bal\'{a}zs Megyeri, and Jon Goldwin$^{*}$}

\address{Midlands Ultracold Atom Research Centre, School of Physics and Astronomy, \\University of Birmingham, Edgbaston, Birmingham B15 2TT, UK}

\email{$^*$j.m.goldwin@bham.ac.uk} %% email address is required

\begin{abstract}
We study electromagnetically induced transparency (EIT) in a heated potassium vapor cell, using a simple optical setup with a single free-running diode laser and an acousto-optic modulator. Despite the fact that the Doppler width is comparable to the ground state hyperfine splitting, transparency windows with deeply sub-natural line widths and large group indices are obtained. A longitudinal magnetic field is used to split the EIT feature and induce magneto-optical anisotropy. Using the beat note between co-propagating coupling and probe beams, we perform a heterodyne measurement of the circular dichroism (and therefore birefringence) of the EIT medium. The observed spectra reveal that lin$\parallel$lin polarizations lead to greater anisotropy than lin$\perp$lin. A simplified analytical model encompassing sixteen Zeeman states and eighteen $\Lambda$ subsytems reproduces the experimental observations. \end{abstract}

 % REPLACE WITH CORRECT OCIS CODES FOR YOUR ARTICLE, MINIMUM OF TWO; Avoid using the OCIS codes for “General” or “General science” whenever possible.

%%%%%%%%%%%%%%%%%%%%%%% References %%%%%%%%%%%%%%%%%%%%%%%%%

%%%%%%%%%%%%%%%%%%%%%%%%%%  body  %%%%%%%%%%%%%%%%%%%%%%%%%%
\section{Introduction}

Electromagnetically induced transparency (EIT) occurs with three-level atoms when a pair of optical transitions simultaneously addressed by a weak probe and strong coupling field share a common atomic state \cite{Boller91}. This dramatically modifies the optical susceptibility of an atomic vapor as a result of quantum interferences between competing scattering channels \cite{Fleischhauer05}. Even though the refractive index in such a gas may differ from unity by less than a part per million, the dispersion line width can be deeply sub-natural, leading to a giant group index of refraction. Critically, this enhanced dispersion is accompanied by reduced absorption. Because of this, and the fact the dispersion is dynamically controllable, EIT and related techniques have generated widespread interest for a number of applications, including slowing and storage of light \cite{Lukin03}, optical magnetometry \cite{Budker02}, and precision timekeeping \cite{Vanier05}.

Here we study EIT in the $\Lambda$ configuration in a heated potassium vapor. In contrast with more commonly used alkali atoms such as cesium and rubidium, EIT in potassium is complicated by the fact that the frequency scale of Doppler broadening is comparable to the ground state hyperfine splitting. This can result in unwanted probe absorption from atoms that are in the wrong combination of internal state and velocity class to undergo EIT. However the relatively small hyperfine splitting in potassium also offers some practical advantages, for example with respect to the generation and detection of a mutually coherent pair of probe and coupling beams. Furthermore in studies of coherent population trapping (CPT) with potassium, it was shown that the overlapping Doppler profiles of the hyperfine ground states lead to reduced optical pumping and therefore enhanced CPT contrast when compared to cesium \cite{Gozzini09}. This work was later extended to EIT in antirelaxation-coated cells \cite{Nasyrov15}. The effects of optical pumping on CPT with cold potassium atoms also have been studied with an eye towards nuclear beta decay experiments with radioactive $^{38{\rm m}}$K and $^{37}$K \cite{Gu03}. The small hyperfine splitting in potassium additionally makes it possible to enter the Paschen-Back regime at significantly lower magnetic fields than in cesium or rubidium \cite{Sargsyan15}. More recently $\Lambda$-EIT-based cooling was used to enable lattice-site-resolved studies of correlated many-body states of quantum degenerate $^{40}$K Fermi gases \cite{Haller15,Edge15}, and ladder-EIT spectroscopy of Rydberg states in heated $^{39}$K has been studied for future experiments with Rydberg-dressed quantum gases of $^{40}$K \cite{Xu16}. In the present work we show how the richness of potassium EIT can be enhanced with the addition of a uniform magnetic field. The resulting Zeeman shifts lift the manifold degeneracies to induce a magneto-optical anisotropy whose polarization dependence highlights the role of coherence.

The rest of the paper is organized as follows. In Section 2 we describe the experimental setup, and investigate the effects of vapor pressure and laser detuning on the group index of the gas. In Section 3 we study the system under the influence of an applied longitudinal magnetic field, and for different coupling beam polarizations. We exploit our configuration of co-propagating and spatially mode-matched probe and coupling beams to demonstrate a heterodyne measurement of the circular dichroism of the gas. We observe a significant dependence of the anisotropy on the angle between linear probe and coupling beam polarizations. A multi-level theoretical model is introduced and used to verify this result. Finally, in Section 4 we discuss our results and conclude.

\section{Absorption coefficient and group index}

\begin{figure}[ht]
\centering\includegraphics[width=\textwidth]{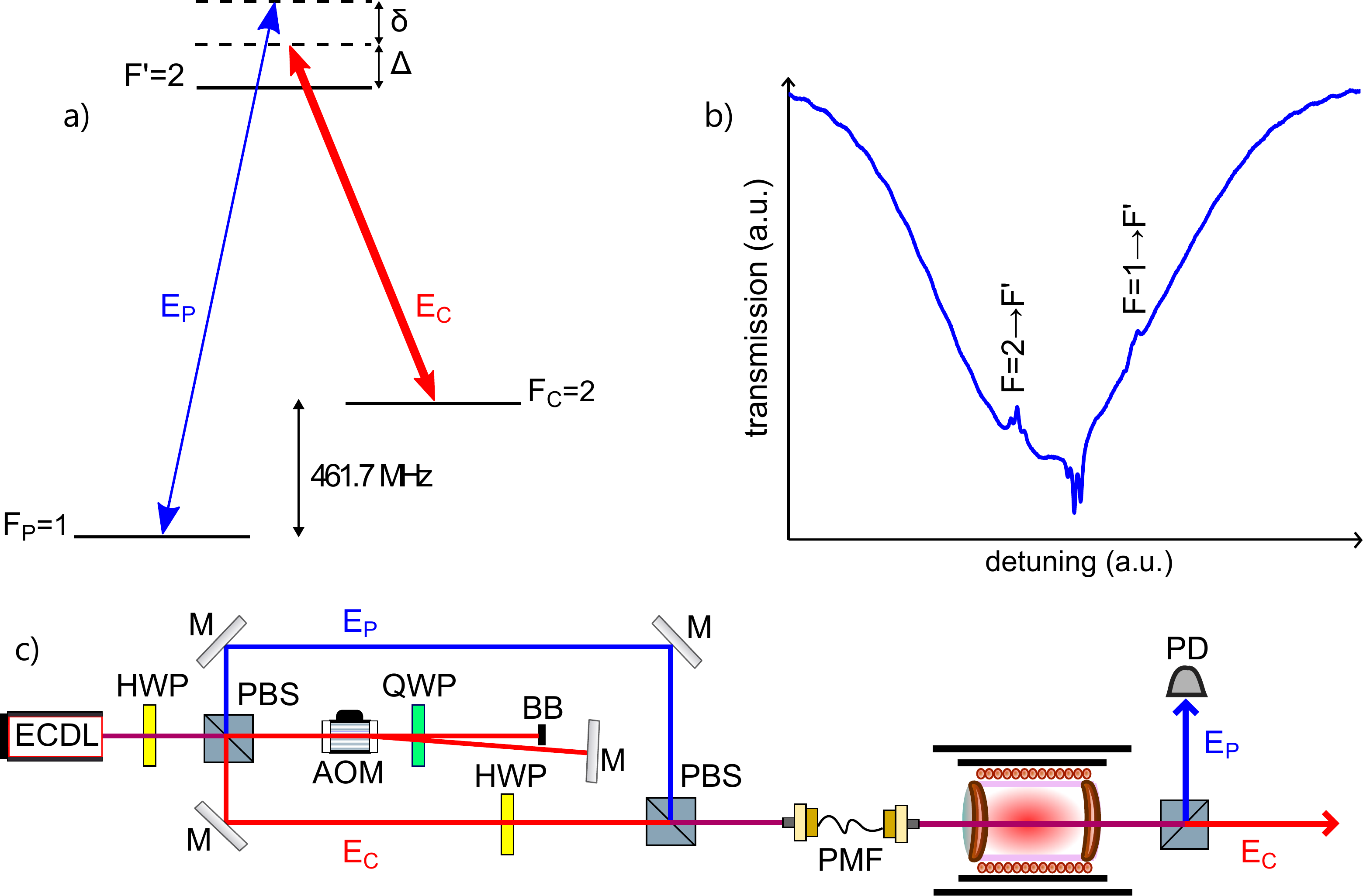}
\caption{(a) Simplified energy schematic of the D$_1$ hyperfine transitions of $^{39}$K used in this work. The coupling beam field $E_c$ (red) is detuned by $\Delta$ from the excited state $F'=2$ manifold ($F'=1$ has been omitted for clarity), and the probe ($E_p$, blue) is detuned by $\delta$ from Raman resonance. The ground state hyperfine splitting is taken from  \cite{Falke06}. (b) Saturated absorption spectrum with no coupling beam present, showing the transitions in (a) and the background Doppler profile. The partially resolved features include excited state crossover resonances, and the transmission dips between manifolds are due to ground state crossovers. (c) Experimental layout, as described in the text. ECDL: external cavity diode laser; M: mirror; HWP: half-wave plate; QWP: quarter-wave plate; PBS: polarizing beam splitter; AOM: acousto-optic modulator; BB: beam block; PMF: polarization-maintaining fiber; PD: photodiode.}
\label{exp_setup}
\end{figure}

We implement EIT on the D$_1$ ($4^2S_{1/2}\leftrightarrow 4^2P_{1/2}$) lines of potassium, shown schematically in Fig.~\ref{exp_setup}(a). The relatively small isotope shifts and hyperfine splittings mean that all of the D$_1$ transitions for the two most naturally abundant potassium isotopes occur within a single Doppler-broadened profile. Given the $93\%$ abundance of $^{39}$K, the $^{41}$K lines are typically not observed in our experiment (see Fig.~\ref{exp_setup}(b)), but the probe and coupling beams still address multiple $^{39}$K transitions. Considering the relative oscillator strengths and our typical detunings, we associate the probe with the $|F_p=1\rangle\leftrightarrow|F'=2\rangle$ manifold of $^{39}$K transitions, and the coupling beam with $| F_c=2\rangle\leftrightarrow|F'=2\rangle$, where $F$ is the total electronic plus nuclear angular momentum and primes denote an excited state.

The experimental setup is outlined in Fig.~\ref{exp_setup}(c). The coupling beam is obtained from an external cavity diode laser running at a wavelength $\lambda=770.1$~nm, and the probe beam, detuned blue by around $461.7$~MHz, is generated by splitting off a small fraction of the laser power and double-passing an acousto-optic modulator (AOM). The two beams are recombined on a polarizing beam splitter (PBS) and coupled into orthogonal axes of a single-mode polarization-maintaining fiber. The beams are collimated together out of the fiber to a $1/e^2$ intensity diameter of $\sim 7.5$~mm. Because the probe and coupling beams are co-propagating, the two-photon EIT transition is effectively Doppler-free; from the ground-state hyperfine splitting, we calculate a broadening of 600~Hz, which is negligible for even our narrowest features. Since the interference effect responsible for EIT is independent of the coupling beam detuning $\Delta$, a large fraction of atoms can contribute to the signal with relatively little dependence on the laser frequency. The relevant scale is set by the single-photon Doppler width $\Delta_D$, which is typically $\sim2\pi\times 400$~MHz in half-width at half-maximum (HWHM). We therefore are able to make all of the measurements described here without any active stabilization of the laser frequency.

The probe and coupling beams co-propagate through a commercial reference cell with an internal length of $L=70$~mm and a natural abundance of potassium, with no buffer gas or anti-relaxation coating. Thin-foil resistive heaters at each end of the cell control the vapor pressure and prevent condensation of potassium on the windows. Although the heater wires trace out a meandering path to reduce stray magnetic fields, we find it necessary to momentarily turn them off during measurements in order to obtain the best transparency features. The assembly is placed within an aluminum lens tube sealed with anti-reflection-coated windows at each end and wrapped in thermal insulation. To control the magnetic field a solenoid is wound around the lens tube and the system is placed in a 24~cm long, two-layer cylindrical mu-metal shield without end caps. After the cell, the probe and coupling beams are separated by another PBS, and the probe power is detected with a biased photodiode whose output current is loaded by the 1~M$\Omega$ input impedance of a digital oscilloscope.

We first demonstrate that large refractive group indices can be obtained, even though the Doppler width approaches the ground state hyperfine splitting. For a probe-frequency-dependent phase index $n(\omega_p)$, the group index $n_g$ is defined as,
\begin{eqnarray}
n_{g}=n+\omega_p\,\frac{{\rm d}n}{{\rm d}\omega_p} \quad.
\label{group_index}
\end{eqnarray}
To obtain $n_g$ from our transmission spectra, we apply the method recently demonstrated with Doppler-broadened media in the absence of EIT ($E_c=0$) \cite{Whittaker15}. As in that work, our EIT medium can be described by a complex susceptibility $\chi(\omega_p)=\chi' + i\chi''$, describing the linear response with respect to the probe field. The real part $\chi'$ describes a phase shift picked up by the probe light as it traverses the cell, with $\chi'=n^2-1$, and the imaginary part $\chi''>0$ describes absorption through the absorption coefficient $\alpha=k\chi''$, where $k=2\pi/\lambda$ is the wavenumber. The fraction of the incident probe power transmitted through the cell is given by,
\begin{eqnarray}
%T(\omega_p) &=& \exp[-kL\chi''(\omega_p)] \label{eq:Tchi} \\
T(\omega_p) &=& \exp[-\alpha(\omega_p)L] \quad. 
\label{eq:Talpha}
\end{eqnarray}
Although linear in $E_p$, the response of our EIT medium is nonlinear in $E_c$, which varies throughout the volume of the cell due to the Gaussian intensity profile and absorption along $z$. Therefore in applying the method from \cite{Whittaker15} to our system, we must understand $\chi$ to represent the susceptibility averaged over the entire cell volume and weighted by the transverse intensity profile, and we assume that the variation in $T(\omega_p)$ around resonance is not too large. Then $\chi''(\omega_p)$ can be calculated from transmission measurements according to Eq.~(\ref{eq:Talpha}). As the medium obeys causality, the real and imaginary parts of $\chi$ are related via the Kramers-Kronig relations; $\chi'(\omega_p)$ can be obtained by applying a Hilbert transform to $\chi''(\omega_p)$, and numerical differentiation of $n(\chi'(\omega_p))$ gives $n_g(\omega_p)$ according to Eq.~(\ref{group_index}).

\begin{figure}[ht]
\centering\includegraphics[width=\textwidth]{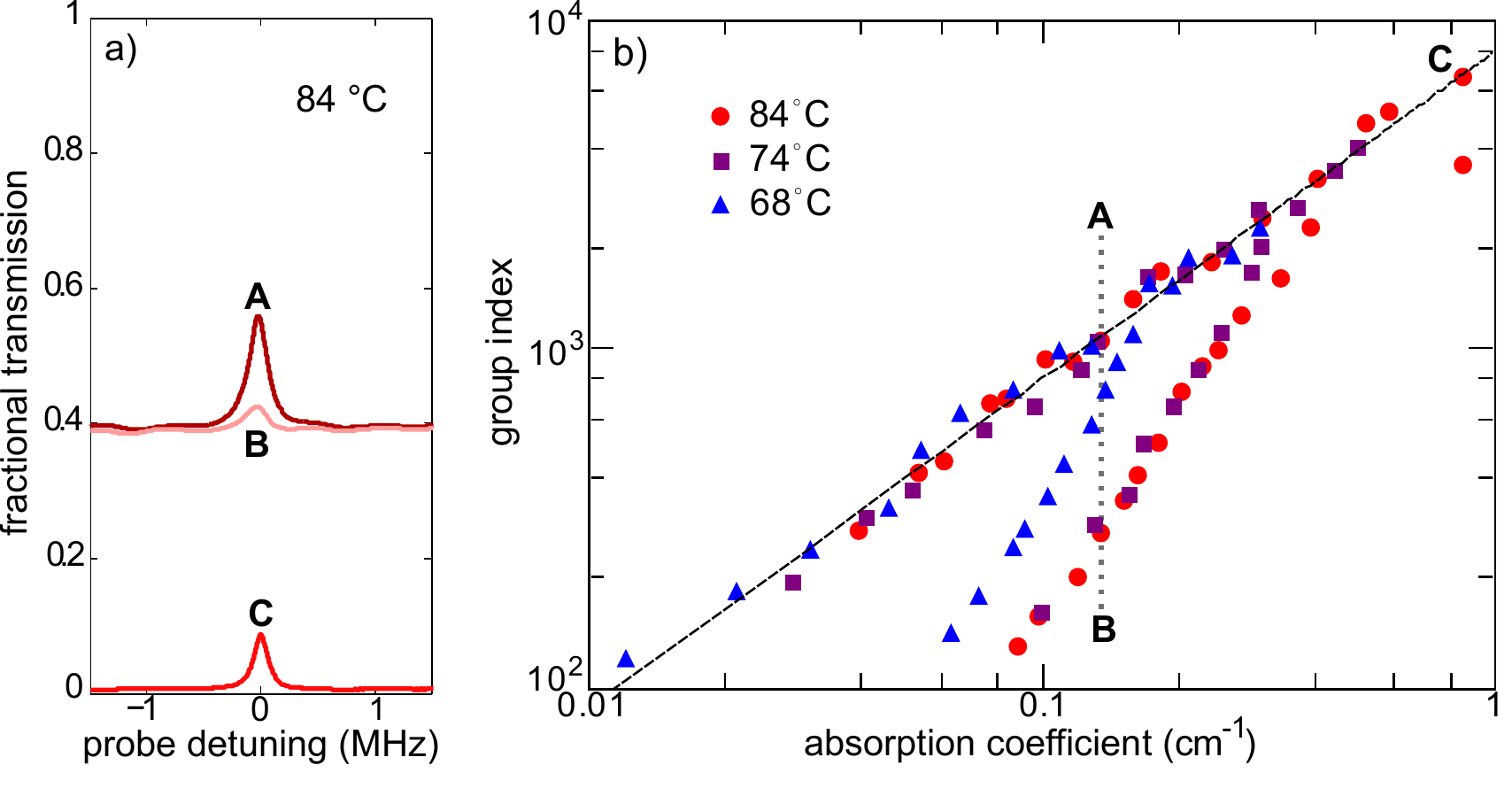}
\caption{Absorption and group index. (a) Fractional transmission spectra $T(\omega_p)$ for coupling (probe) power of $300~(50)~\mu$W and a cell temperature of $84^\circ$C. The background absorption coefficient is varied by coarse tuning of the laser, as described in the text. The EIT linewidth is $2\pi\times\{96,98,72\}$~kHz for trace \{A,B,C\}, and the amplitude is $\{16,3,8\}\%$. (b) Effect of optical depth on the group index. For each cell temperature, the group index is measured for varying laser detuning $\Delta$, leading to varying background probe absorption (i.e., just outside the EIT resonance). The upper branches correspond to probing on the blue side of the Doppler profile near the $F_p\to F'$ transitions, and the bottom halves on the red side near $F_c\to F'$. The solid line shows the linear dependence expected for simple EIT.}
\label{fig:ng}
\end{figure}

Example transmission spectra are shown in Fig.~\ref{fig:ng}(a) for a cell temperature of $84^\circ$C and a selection of laser detunings. The Raman detuning $\delta$ is swept using a voltage-controlled oscillator driving the AOM. The coupling beam power at the entrance to the cell is $300~\mu$W and the probe beam is $50~\mu$W. In trace C the laser is tuned so that the probe is near the center of the combined single-photon Doppler profile shown in Fig.~\ref{exp_setup}(b), giving the strongest background absorption. The fractional transmission in C varies by $8\%$ over a frequency range of $2\pi\times 72$~kHz (HWHM), which is much smaller than the corresponding natural line width, $\gamma=2\pi\times 2.978$~MHz \cite{Wang97}. For these spectra the EIT line width is dominated by power broadening from the coupling beam. The incident power of $300~\mu$W was chosen to maximise the group index in a compromise between increasing EIT contrast and broadening with increasing power. In other measurements (not shown) we have observed transparency windows below $2\pi\times 50$~kHz ($=\gamma/60$), with a linear dependence on coupling beam power and a y-intercept equal to $2\pi\times(43.0\pm1.0)$~kHz. Of this limiting value we calculate a contribution of $2\pi\times 11$~kHz from transit broadening, and a lower limit of $2\pi\times 17$~kHz from $B$-field variations of $1.2~\mu$T throughout the interaction region, as calculated on axis for our relatively small solenoid.

As mentioned previously, because of our co-propagating probe and coupling beams, a transparency feature is observed around zero Raman detuning ($\delta=0$) independent of the coupling beam detuning $\Delta$. Traces A and B in Fig.~\ref{fig:ng}(a) are taken with the probe beam on the blue- and red-detuned sides, respectively, of the Doppler profile. Although the spectra were chosen to have similar $\alpha$ outside of the EIT feature, the heights of the peaks are significantly different. In A the probe and coupling beams are near resonance with the transitions shown in Fig.~\ref{exp_setup}(a) in the lab frame, leading to optimum EIT. In contrast, in B the probe absorption is dominated by high-velocity atoms in the wrong ($F=2$) ground state, which see the probe beam red-shifted into resonance. For these atoms a $\Lambda$ system exists with the coupling beam addressing the $(F=1)\leftrightarrow F'$ transitions, but with Raman detuning increased by the ground state hyperfine splitting, so that transparency is no longer induced. The resulting excess probe absorption leaves only a small EIT feature corresponding to the $\Lambda$ systems shown in Fig.~\ref{exp_setup}(a), but with very large $\Delta$. If off-resonant absorption is negligible, the group index should be proportional to the absorption coefficient. We measured $n_g$ for varying laser detuning at multiple cell temperatures, a sample of which are shown in Fig.~\ref{fig:ng}(b). The upper branch of each data set corresponds to probe detunings on the blue side of the combined Doppler profile. In this regime $n_g$ is proportional to $\alpha$ for all of the temperatures, as shown by the linear fit (dashed line). On the red side the group index is much smaller for a similar absorption coefficient, showing the effects of single-photon absorption. The peak group index monotonically increases with cell temperature until $\sim 90^\circ$C. Beyond this, the maximum achievable $n_g$ is reduced due to excessive absorption of the coupling beam. The cell was therefore kept at $84^\circ$C for the rest of the experiments described here. These results imply that under optimum conditions, a light pulse whose frequency bandwidth is below the EIT line width will propagate through the cell with a group velocity $v_g=c/n_g$ more than 6000 times slower than the vacuum speed of light, $c$.

\section{Magneto-optical anisotropy}

When applying a longitudinal magnetic field, the EIT resonance is split into three distinct components. Such splitting of EIT and CPT features was the subject of early studies \cite{Schmidt96,Wynands98} and interest in this phenomenon has continued \cite{Wei05,Iftiquar09,Sargsyan15}. The resonance frequency of the central peak is independent of field to first order, while the side peaks are shifted with effective magnetic moments equal to $\pm1$ Bohr magneton ($\mu_B=2\pi\hbar\times 14~$kHz/$\mu$T). This occurs because the linear Zeeman shift perturbs the ground state energies by $g_Fm_F\mu_BB$ where $m_F$ is the projection of $F$, and the Land\'{e} $g$-factor $g_F$ equals $+1/2$ for the $F_c$ manifold and $-1/2$ for $F_p$ ($g_{F'}=\pm 1/6$, but shifts to the excited states do not affect the Raman resonance frequencies). For linear polarizations, both probe and coupling fields can be decomposed along the $B$-field axis into superpositions of left- and right-handed circular polarizations, so that every $\Lambda$ system contributing to EIT connects pairs of ground states with $\Delta m_F=0$ or $\pm2$, giving frequency shifts of 0 or $\pm\mu_BB/\hbar$. The central peak shows no anisotropy. Although the side peaks are not uniquely associated with either $\sigma^\pm$ atomic transitions, they do exhibit circular dichroism and birefringence. The latter manifests itself as Faraday rotation of a linearly polarized probe, which can be detected using a balanced polarimeter \cite{Budker02}. Coherent control of polarization rotation in $\Lambda$ systems with and without magnetic fields has been studied in cold lithium \cite{Franke01}, cesium \cite{Choi07}, and rubidium \cite{Wojciechowski10}, and in heated rubidium \cite{Wang06} and sodium \cite{Hombo12}.

\begin{figure}[ht]
\centering\includegraphics[width=8cm]{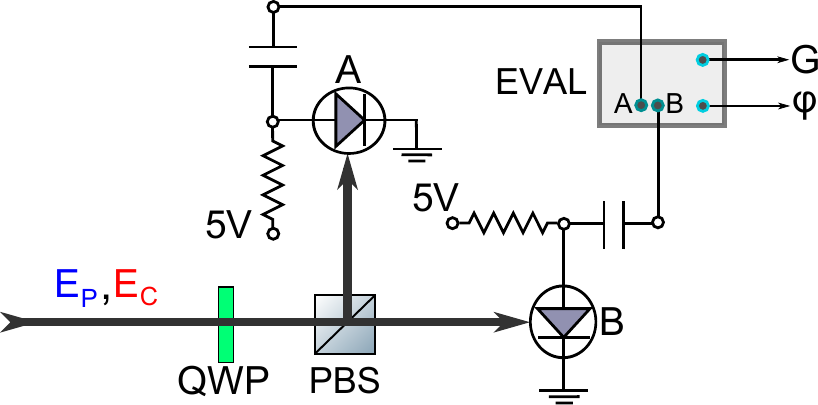}
\caption{Simplified schematic for heterodyne measurements of optical anisotropy. The quarter-wave plate and PBS act together as a circularly polarizing beam splitter according to Eq.~(\ref{eq:QWP}). The probe and coupling beam components interfere at the fast photodiodes A and B, and the resulting beat notes are used as inputs to a gain/phase detecting circuit (EVAL) as described in the text.}
\label{fig:dich_setup}
\end{figure}

In $\Lambda$ systems, where the probe and coupling beam wavelengths are similar, polarization rotation measurements typically require some relative misalignment of the probe and coupling beams, to prevent the latter from entering the polarimeter. We prefer not to sacrifice our compact, robust, and Doppler-free configuration. Instead we have developed a measurement of magneto-optical activity which exploits the spatial mode-matching of our beams. The setup is shown schematically in Fig.~\ref{fig:dich_setup}. A quarter-wave plate (QWP), with its fast axis at $45^\circ$ from horizontal, converts right- and left-hand circular polarizations to vertical and horizontal. In the language of Jones vectors \cite{Steck},
\begin{eqnarray}
\left.\begin{array}{c}
{\mathbf e}_r \\
{\mathbf e}_l	
\end{array}\right\} &\stackrel{\rm QWP}{\longrightarrow}& \left\{
\begin{array}{l}
{\mathbf e}_v\,e^{-i\pi/4} \\
{\mathbf e}_h\,e^{i\pi/4}
\end{array}\right. \quad,
\label{eq:QWP}
\end{eqnarray}
where ${\mathbf e}_r$ and ${\mathbf e}_l$ are the basis vectors for right- and left-handed circular polarizations (RHC and LHC, respectively), which are related to horizontal ($h$) and vertical ($v$) polarizations by ${\mathbf e}_r=({\mathbf e}_h-i{\mathbf e}_v)/\sqrt{2}$ and ${\mathbf e}_l=({\mathbf e}_h+i{\mathbf e}_v)/\sqrt{2}$.

After the wave plate the horizontal and vertical components of both probe and coupling beams are separated on a polarizing beam splitter and sent to two high-speed photodetectors (Hamamatsu G4176-03 \cite{products}), which are biased at $+5$~V with bias-tees (Mini-Circuits ZFBT-4R2GW+). The two microwave beat notes are amplified by 24~dB (Mini-Circuits ZFL-500LN+), and then sent to the inputs of an integrated chip gain/phase detector (Analog Devices AD8302-EVALZ), whose outputs are described below.

To calculate the detected signal, we first consider the propagation of the probe beam. The incident field is assumed to take the form ${\mathbf E}_p={\mathbf e}_h\,E_p\exp(-i\omega_pt)$. After the cell,
\begin{eqnarray*}
{\mathbf E}_p(\mbox{after cell}) &=& \frac{E_p\,e^{-i\omega_pt}}{\sqrt{2}}\,\left({\mathbf e}_r\,e^{-\alpha_rL/2}e^{i\phi_r} + {\mathbf e}_l\,e^{-\alpha_lL/2}e^{i\phi_l} \right) \quad,
\end{eqnarray*}
where $\alpha_i$ is the absorption coefficient for $i$-handed circular polarization, and the corresponding phase shift $\phi_i$ is related to the phase index by $\phi_i=kLn_i$. After the quarter-wave plate,
\begin{eqnarray}
{\mathbf E}_p(\mbox{after QWP}) &=& \frac{E_p\,e^{-i\omega_pt}}{\sqrt{2}}\,\left[\,{\mathbf e}_v\,e^{-\alpha_rL/2}e^{i(\phi_r-\pi/4)} + {\mathbf e}_h\,e^{-\alpha_lL/2}e^{i(\phi_l+\pi/4)} \right] \quad. \label{eq:Ep}
\end{eqnarray}
The coupling beam is initially of the form ${\mathbf E}_c={\mathbf e}_j\,E_c\exp(-i\omega_ct)$, where $j=v$ for the lin$\perp$lin configuration coming from the fiber, and $j=h$ for lin$\parallel$lin, obtained by placing a polarizer at $45^\circ$ before the cell and increasing the incident power to obtain the same power immediately before the cell. Given the relative probe and coupling beam powers, we assume the coupling beam absorption coefficient $\alpha_c$ and phase $\phi_c$ are both independent of $\delta$ for a fixed value of $\Delta$, and take $\phi_c=0$ without loss of generality. Then,
\begin{eqnarray}
{\mathbf E}_c(\mbox{after QWP}) &=& \frac{E_c\,e^{-i\omega_ct}}{\sqrt{2}}\,\left[\,{\mathbf e}_v\,e^{-\alpha_cL/2}e^{\mp i\pi/4} + {\mathbf e}_h\,e^{-\alpha_cL/2}e^{\pm i\pi/4} \right] \quad, \label{eq:Ec}
\end{eqnarray}
where the upper and lower symbols refer to lin$\parallel$lin and lin$\perp$lin configurations, respectively. After the polarizing beam splitter, the detected beat notes are proportional to $2{\rm Re}({\mathbf E}_p\cdot{\mathbf E}_c^\ast)$,
\begin{eqnarray}
\mbox{LHC ($h$ after PBS)} &=& E_p E_c e^{-(\alpha_l+\alpha_c)L/2}\begin{array}{c}\cos \\[-0.15cm] \sin \end{array}\!\!\!(\delta\!\omega\,t-\phi_l) \\
\mbox{RHC ($v$ after PBS)} &=& \pm E_p E_c e^{-(\alpha_r+\alpha_c)L/2}\begin{array}{c}\cos \\[-0.15cm] \sin \end{array}\!\!\!(\delta\!\omega\,t-\phi_r) \quad,
\end{eqnarray}
with the beat note frequency $\delta\!\omega=\omega_p-\omega_c$. The AC signals from the two photodiodes are sent to the A and B input channels of the evaluation board. The board generates two voltage outputs: one linear in the logarithm of the ratio of A/B amplitudes, and one linear in the phase difference. Specifically,
\begin{eqnarray}
\mbox{amplitude ratio:}\quad V_G &=& 0.9~{\rm V}-0.6~{\rm V}\left(\frac{\Delta\alpha\,L}{2\ln 10}\right) \label{eq:VG}\\ 
\mbox{phase difference:}\quad V_\varphi &=& 0.9~{\rm V}-1.8~{\rm V}\left(\frac{|\Delta\phi|-\pi/2}{\pi}\right) \quad. \label{eq:Vphi}
\end{eqnarray}
Here $\Delta\alpha=\alpha_r-\alpha_l$ characterizes the circular dichroism and $\Delta\phi=\phi_r-\phi_l$ the birefringence. For simplicity we have dropped a $\pi$ phase shift in the lin$\perp$lin case, which doesn't affect the principle of the measurement. Since the detector expects phase differences centered around $\pi/2$, it is necessary anyway to add a bias phase (we used quarter-wave lengths of cables).

\begin{figure}[ht]
\centering\includegraphics[width=\textwidth]{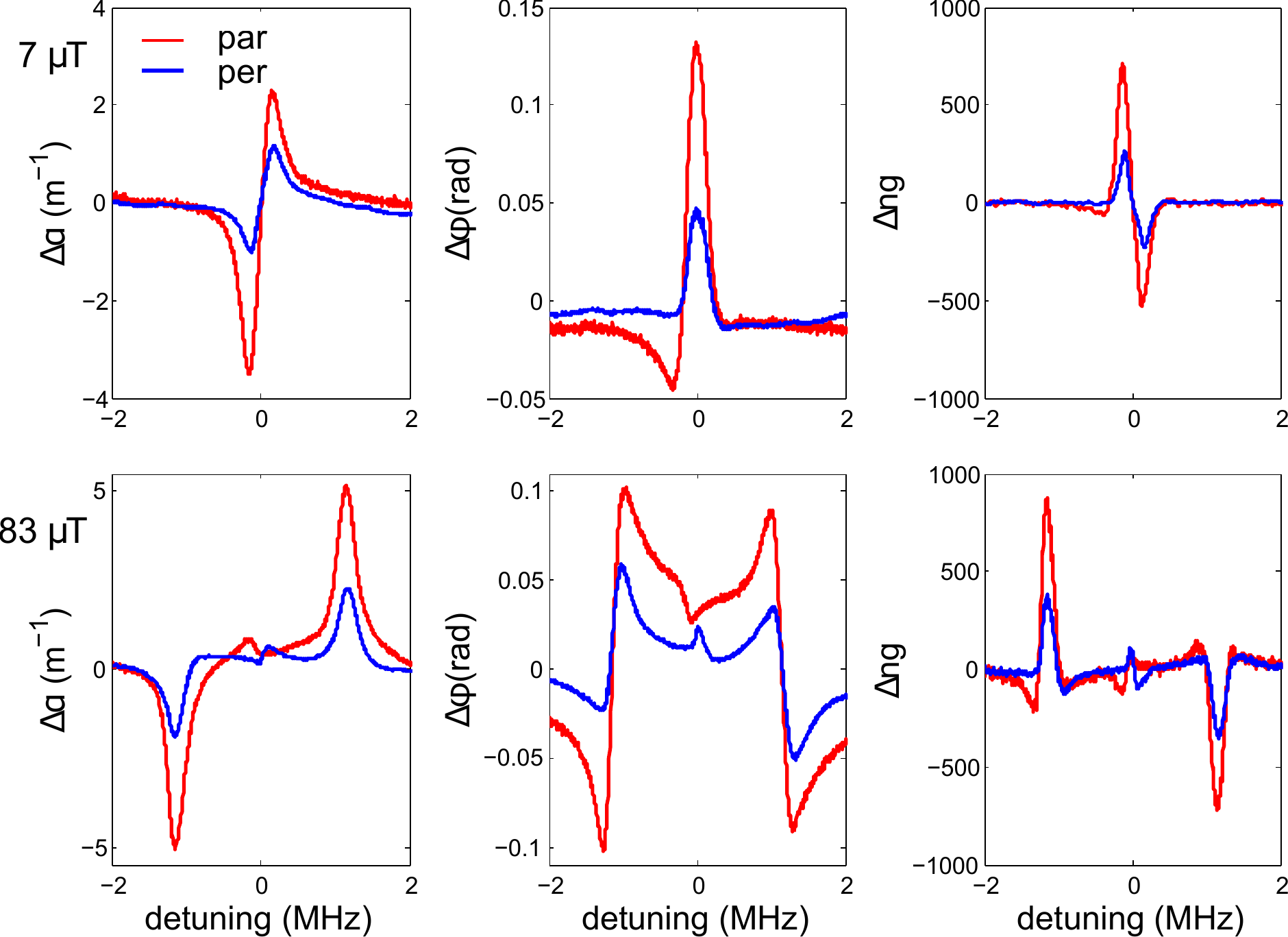}
\caption{Heterodyne spectra of magneto-optically induced anisotropy. Red (blue) traces show the results for lin$\parallel$lin (lin$\perp$lin) polarizations; $\Delta\alpha$ is the differential absorption coefficient, $\Delta\phi$ is the phase difference (obtained by Hilbert transformation of $\Delta\alpha$), and $\Delta n_g$ is the group index difference obtained by numerical differentiation of $\Delta\phi$ after smoothing with a cubic Savitzky-Golay filter (implemented in Matlab using the {\it sgolayfilt} command).}
\label{fig:spectra}
\end{figure}

Equations (\ref{eq:VG}) and (\ref{eq:Vphi}) show that our method produces voltages linear in the differential absorption and phase shift between circular polarization components, without the need to normalize to the full transmission. Furthermore, since the probe and coupling beams propagate together, there is strong common-mode rejection of any phase noise due to mechanical vibrations of the optics. Example spectra are shown in Fig.~\ref{fig:spectra}. The detuning $\Delta$ and coupling beam power (1~mW) were adjusted to maximize the dichroism at large fields with $50~\mu$W of probe light. We found that our values of $\Delta\phi$ were too small to be measured directly given the higher noise level of the phase channel output. Therefore in practice we obtained spectra of $\Delta\phi$ via Hilbert transforms of $\Delta\alpha$ in the spirit of what has been described previously. 

We observe that the magneto-optical anisotropy of the EIT medium is systematically 2--3 times greater for lin$\parallel$lin polarizations than for lin$\perp$lin. The dichroism $\Delta\alpha$ is of order $\alpha/10$ for lin$\parallel$lin polarizations. This is accompanied by an equivalent polarization rotation angle $\theta=\Delta\phi/2\sim 70$~mrad, giving a Verdet constant of $\mathcal{V}\sim 3\times 10^5$~rad/T/m for small fields. The group index difference implies a relative delay of $\Delta t=\Delta n_g L/c\sim 200$~ns between LHC and RHC probe components. 

The observed difference between polarization configurations can be attributed to magneto-optical coherence. The linearly polarized probe and coupling beams each represent 50/50 superpositions of LHC and RHC light, with only the relative phases of the components differing between lin$\parallel$lin and lin$\perp$lin. These phases can influence the anisotropy only if the probability amplitudes describing scattering within individual $\Lambda$ subsystems add coherently. It is not {\it a priori} clear whether this should be the case. Schmidt {\it et al.}~studied the $B$-field splitting of CPT resonances in \cite{Schmidt96}, and found in that case that the subsystems summed incoherently to the total signal. However EIT experiments without magnetic fields have shown some unexpected and non-trivial results attributed to coherences between multiple scattering channels \cite{Chen00,Li09,Mishina11}. Our results similarly show a clear dependence on the relative phase of LHC and RHC polarizations, highlighting the role of coherence in generating the magneto-optical anisotropy. To the best of our knowledge this specific effect has not been observed before.

To model our EIT medium, we begin with the susceptibility for a single $\Lambda$ system, obtained from the three-level master equation \cite{Fleischhauer05},
\begin{eqnarray}\label{eq:chi3LA}
\chi &=& i\,\frac{\mathcal{N}d_p^2}{\hbar\varepsilon_0}\frac{\tfrac{1}{2}\Gamma-i\delta}{|\tfrac{1}{2}\Omega_c|^2+(\tfrac{1}{2}\Gamma-i\delta)[\gamma-i(\delta+\Delta)]} \quad.
\end{eqnarray}
Here $\mathcal{N}$ is the number density of atoms, $d_p$ is the atomic dipole moment for the probe transition, and $\Omega_c=-{\bf d}_c\cdot{\bf E}_c/\hbar$ is the Rabi frequency associated with the coupling beam. A ground state dephasing rate $\Gamma$ has been added to account for broadening of the EIT features due to magnetic field variations and the finite transit time of atoms passing through the beams.

With the quantization axis taken along the $B$-field and propagation direction, the linearly polarized probe and coupling beams comprise a total of four circularly polarized components. If we include the $F'=1$ excited states, which are well within the Doppler-broadened profile, then we have a total of 18 interconnected $\Lambda$ systems encompassing 16 Zeeman states, each defined by the transitions $|F_p,m_p\rangle\stackrel{p}{\longleftrightarrow}|F',m_p+q_p\rangle\stackrel{c}{\longleftrightarrow}|F_c,m_p+q_p-q_c\rangle$, where the $q=\pm1$ correspond to spherical tensor components of the probe and coupling fields. Rather than solve the full 18-level master equation, we simply treat each $\Lambda$ system individually according to Eq.~(\ref{eq:chi3LA}), with $d_p$ and $\Omega_c$ depending on the relevant oscillator strengths and coupling beam polarization component, and with modified detunings,
\begin{eqnarray}
\delta &\to& \delta + \mu_B B\,[\,g_{F_p}m_p-g_{F_c}(m_p+q_p-q_c)] \\
(\delta+\Delta) &\to& (\delta+\Delta) + \Delta_{F'} + \mu_B B\, [\,g_{F_p}m_p-g_{F'}(m_p+q_p)] \quad.
\end{eqnarray}
The additional excited state detuning $\Delta_{F'}$ is taken to be $2\pi\times 55.55$~MHz for $F'=1$, and 0~MHz for $F'=2$ \cite{Falke06}, and the distribution of atomic velocities are accounted for by averaging over Doppler shifted values of $\Delta$ weighted by a Maxwell-Boltzmann distribution. Summing over $m_p$ and $F'$ for fixed $q_p$ and $q_c$ gives a single component $\chi_{q_p}^{q_c}$, and then the total $\chi_{q_p}$ is given by a symmetric (anti-symmetric) superposition of $\chi_{q_p}^{+1}$ and $\chi_{q_p}^{-1}$ for lin$\parallel$lin (lin$\perp$lin) polarizations. We thus finally obtain the susceptibility difference,
\begin{eqnarray}\label{eq:Deltachi}
\nonumber
\Delta\chi &=& \chi_{-1}-\chi_{+1} \\ 
&=& \left\{
\begin{array}{cl}
 \big(\chi_{-1}^{+1}+\chi_{-1}^{-1}\big) - \big(\chi_{+1}^{-1}+\chi_{+1}^{+1}\big), & \mbox{lin}\parallel\mbox{lin} \\[0.6em]
 \big(\chi_{-1}^{+1}-\chi_{-1}^{-1}\big) - \big(\chi_{+1}^{-1}-\chi_{+1}^{+1}\big), & \mbox{lin}\perp\mbox{lin}
\end{array} \right.\quad.
\end{eqnarray}

\begin{figure}[h!]
\centering\includegraphics[width=\textwidth,trim={0cm 0cm 0cm 0cm},clip]{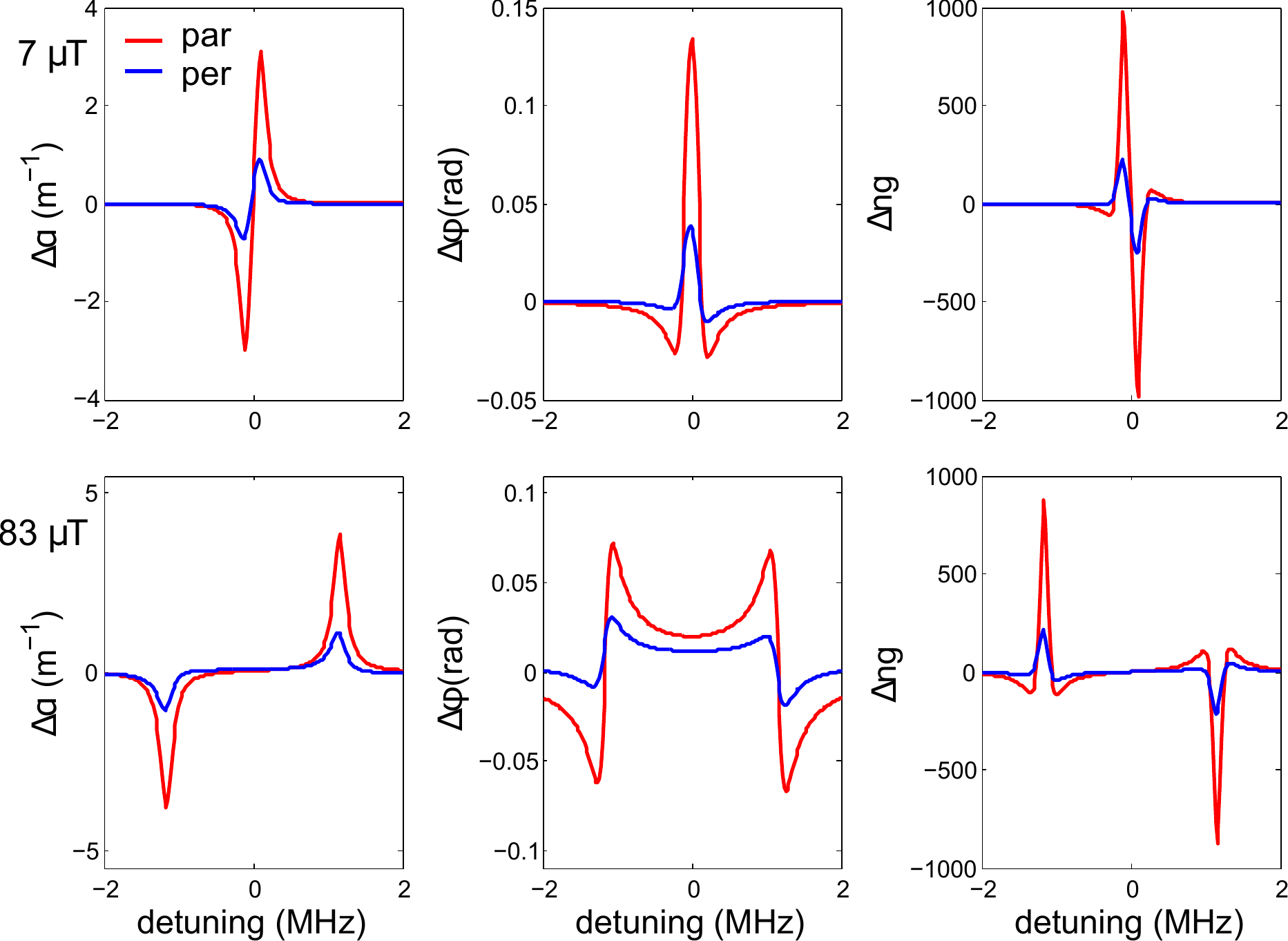}
\caption{Theoretical spectra, calculated using Eqs.~(\ref{eq:chi3LA})--(\ref{eq:Deltachi}). Parameters are $\Delta=0$, $\Omega_c=2\pi\times 10$~MHz, $\Gamma=2\pi\times 0.18$~MHz, and $\mathcal{N}=1.5\times 10^{11}$~cm$^{-3}$ (we expect $1.8\times 10^{11}$~cm$^{-3}$ for our temperature \cite{Hanley15}).}
\label{fig:spectheory}
\end{figure}

The calculated spectra are shown in Fig.~\ref{fig:spectheory}. The main features of the data in Fig.~\ref{fig:spectra} are reproduced, in particular the greater optical anisotropy for lin$\parallel$lin polarizations. One can also see the spectral asymmetry of $\Delta\phi$ due to the additional detuning $\Delta_{F'}$ of the $F'=1$ states. Missing in the simulation is the small remnant of the central EIT peak visible at higher field. This feature only vanishes under perfect balancing of the ideally equal LHC and RHC contributions. Any imperfections in the polarizations (such as induced by birefringence in the heated cell windows) or in the uniformity of the magnetic field (due to the small solenoid and imperfect shielding) will perturb this balance. There are also some quantitative differences in the magnitudes and widths of the features. Otherwise the excellent agreement between this simplified model and our observations suggests that a number of other effects not accounted for in our calculation are negligible for our parameters. For example we have not included the variation in $E_p$ and $E_c$ throughout the cell volume, have implicitly assumed a uniform distribution of ground state populations, and have neglected nonlinearities and higher-order coherences related to the multiply-connected network of $\Lambda$ systems. Still the magnitude, spectral shape, and polarization dependence of the magneto-optical anisotropy are all remarkably well modeled by Eqs.~(\ref{eq:chi3LA})--(\ref{eq:Deltachi}). We stress that all of the $\chi_{q_p}^{q_c}$ terms in Eq.~(\ref{eq:Deltachi}) are the same for the two polarization configurations, and it is only the signs of their contributions which differ. These signs have their origins in the superpositions of LHC and RHC polarizations in the probe and coupling beams, revealing the coherent nature of the magneto-optical anisotropy.

\section{Discussion and conclusion}

We have studied electromagnetically induced transparency in a heated potassium vapor cell. Features with deeply sub-natural line widths, limited by transit broadening and magnetic field inhomogeneities, were obtained with only a free-running laser and an acousto-optic modulator. Optimization of the group index with cell temperature resulted in group indices of several thousand, and the effect of absorption from Doppler-shifted atoms not participating in EIT was observed. The response of the medium to a longitudinal magnetic field was studied, showing that the EIT medium exhibits magneto-optical anisotropy for either lin$\perp$lin or lin$\parallel$lin probe and coupling beam polarizations. A novel heterodyne measurement of the circular dichroism was demonstrated, and used to show that the magnitude of the induced anisotropy is sensitive to the relative polarizations. This effect highlights the role of interfering coherences among the many $\Lambda$ systems contributing to the signal, as captured in a microscopic multilevel model. 

Our longer-term interest is in applying these methods to laser-cooled potassium clouds, where we can expect much less residual absorption at EIT resonance, and narrower line widths. In our main experiment $^{39}$K atoms are being magneto-optically cooled and trapped within the mode volume of a high-finesse ring cavity. This will enable studies of strong in-cavity dispersion \cite{Wang00,Pati07,Laupretre11,Laupretre12} and lasing with extremely large \cite{Bohnet12,Weiner12} or near-zero \cite{Shahriar07} group indices, for applications in active dispersion-enhanced metrology and sensing.

\section*{Acknowledgments}

We gratefully acknowledge funding from the UK EPSRC (EP/J016985/1) and DSTL (DSTLX1000092132). We thank M.~Holynski for early contributions to the laser system, M.~Colclough and M.~Brannan for the loan of the mu-metal shielding, and A.~Kowalczyk for pointing out the adverse effect of the heaters. We are also grateful to P.~G.~Petrov and V.~Boyer for numerous discussions and feedback on the manuscript. Data presented here are available online \cite{data}.


\begin{thebibliography}{99}
\bibitem{products} Trade names and part numbers are used for identification purposes only, and do not constitute endorsements. Other products may perform similarly or better.
\end{thebibliography}
\end{document}